\documentclass[fleqn,usenatbib]{mnras}

\usepackage[T1]{fontenc}
\usepackage{ulem}
\DeclareRobustCommand{\VAN}[3]{#2}
\let\VANthebibliography\thebibliography
\def\thebibliography{\DeclareRobustCommand{\VAN}[3]{##3}\VANthebibliography}

\usepackage{graphicx}	
\usepackage{amsmath}	
\usepackage{amssymb}	
\usepackage{newtxtext,newtxmath}
\usepackage{xcolor}
\usepackage{float}
\usepackage{bm}
\usepackage{tabularx}
\usepackage{color,soul}

\newcommand{\mkn}[1]{{\textcolor{teal}{#1}}}

\title[ Neutrino Cosmology and Cosmic Web]{Clustering of dark matter in the cosmic web as a probe of massive neutrinos}

\author[Khoshtinat et al.]{
Mohadese Khoshtinat$^{1}$, 
Mohammad Ansarifard$^{2, 5}$,
Farbod Hassani$^{3}$,
Shant Baghram$^{1, 4}$ \thanks{baghram@sharif.edu}
\\ \\
$^{1}$
Department of Physics, Sharif University of
Technology, P.~O.~Box 11155-9161, Tehran, Iran\\
$^{2}$
School of Astronomy, Institute for Research in Fundamental Sciences (IPM), P.O. Box 19395-5531,
Tehran, Iran
\\
$^{3}$
Institute of Theoretical Astrophysics, Universitetet i Oslo, 0315 Oslo, Norway \\
$^{4}$
Research Center for High Energy Physics, Department of Physics, Sharif University of Technology, P.O.Box 11155-9161, Tehran, Iran\\
$^{5}$
Department of Physics, K.N. Toosi University of Technology, P.O. Box 15875-4416, Tehran, Iran\\
}

\date{Accepted XXX. Received YYY; in original form ZZZ}

\pubyear{2024}

\begin{document}
\label{firstpage}
\pagerange{\pageref{firstpage}--\pageref{lastpage}}
\maketitle
\begin{abstract}
The large-scale structure of the Universe is distributed in a cosmic web. Studying the distribution and clustering of dark matter particles and halos 
may open up a new horizon for studying the physics of the dark Universe. In this work, we investigate the nearest neighbour statistics and spherical contact function in cosmological models with massive neutrinos.
For this task, we use the relativistic N-body code, gevolution and study particle snapshots at three different redshifts. In each snapshot, we find the halos and evaluate the letter functions for them. We show that a generic behaviour can be found in the nearest neighbour, $G(r)$, and spherical contact functions, $F(r)$, which makes these statistics promising tools to constrain the total neutrino mass.
\end{abstract}

\begin{keywords}
(cosmology:) dark matter, (cosmology:) large-scale structure of Universe
\end{keywords}



\section{Introduction}
The data sets from ongoing \citep{DESI:2016fyo} and upcoming \citep{Amendola:2016saw} large scale structure (LSS) surveys not only let tighter constraints on the  cosmological parameters but make it possible to address some fundamental questions in physics and cosmology- e.g., the equation-of-state of dark energy, the characteristics of dark matter, and the total neutrino mass. 
With a wealth of raw data, the task is to extract as much possible information in a computationally efficient way from these data sets. Due to the presence of non-linearities, especially in the late-time Universe, this will be a challenging task. 
On one hand, the linear perturbation theory (LPT) still holds on large scales \citep{Dodelson:2003ft}, where the density contrast field can also be considered a Gaussian random field. So, the traditional 2-point correlation function contains all the information and is the most suitable statistic for extracting the information from the LSS data \citep{Romanello:2023obk}. 
On the contrary, at small scales \citep{Cooray:2002dia}, the nonlinear gravitational evolution of matter leads to a non-Gaussian field as a result of mode couplings. In this situation, information diffuses into the higher-order correlation functions, and the 2-point correlation function does not contain all the information. So it fails to grasp the complete content of the data set.
The direct way to capture leaked information from the second moment is to compute higher-order correlation functions, e.g., bispectrum \citep{Sefusatti:2015aex}. 
As going to deep non-linear regimes, it is not straightforward that the information is stored in low moments of the correlation function. It means that the whole probability distribution function (PDF) incorporates the non-Gaussian effects. 
One approach to circumvent these difficulties is to consider the 2-point correlation function (or its Fourier counterpart, the power spectrum) equipped with extra information - e.g., the marked power spectrum \citep{stoyan1984correlations} - or by calculating it in different environments like voids, nodes, sheets and filaments in real space \citep{Bonnaire:2021sie}, or in redshift space \citep{Bonnaire:2022ocm}.
Parallel to these efforts, many non-linear semi-analytical models have been introduced, that use the relation between linear scale matter distribution statistics and non-linear observables such as the number density of dark matter halos. Peak theory \citep{Bardeen:1985tr,Sheth:1999mn} and the excursion set theory \citep{Bond:1990iw,Sheth:1999su, Zentner:2006vw, Nikakhtar:2016bju,Nikakhtar:2018qqg} are two non-linear structure formation models, which use the relation between linear and non-linear scales to probe the distribution of matter over a wide range of scales. 
\\
Another strategy is to search for other statistical measures sensitive to all higher-order correlation functions that are computed for one point in space. To mention a few - the PDF of the matter density field smoothed at some radius \citep{Uhlemann:2019gni,Boyle:2020bqn,Gough:2021hlr}, k-nearest neighbours cumulative distribution functions (CDF) or peaked-CDF \citep{Banerjee:2020umh,Banerjee:2021cmi,Yuan:2023llf}, and the pioneering work of \cite{White:1979kp}  using the void probability function to study the clustering of the tracers, e.g., halos or galaxies. 
\cite{Fard:2021qaa} employed the nearest neighbour CDF, or $G(r)$ and the spherical contact CDF, or $F(r)$ within the context of the standard model of cosmology based on cosmological constant $\Lambda$ and cold dark matter (CDM) known as $\Lambda$CDM model using the SMDPL simulations \citep{Moline:2021rza}  with different mass cuts and at redshifts $z=0,0.5,1$. They demonstrated that these statistical measures could differentiate between samples constructed by mass cuts at various redshifts.

\textit{The subsequent query is if these statistics (i.e., $G$ and $F$) can distinguish between different extensions of the} $\Lambda$CDM. \\
In this paper, we compute these summary statistics for the base $\Lambda$CDM cosmology and cosmologies with massive neutrinos, $\nu\Lambda$CDM.
After a brief review of the impact of massive neutrino on the late time cosmology and the statistics used in this work in Section \ref{Sec2}, we describe the simulations and the data preparation in Section \ref{Sec3}. The results of the analysis are depicted in Section \ref{Sec4}. Finally, besides wrapping up, we point out the future direction in Section \ref{Sec5}.


\section{Theoretical background}\label{Sec2}
In this section, we present the theoretical background needed for this work. In the first subsection, we review the idea of one-point statistics, and in the second subsection, we address the effect of massive neutrinos on large-scale structures.

\subsection{One point statistics}
In the spatial point analysis, we use the cumulative distribution functions of the nearest neighbour distances (NNDs) such as $F(R)$ and $G(R)$ to reveal the clustering feature of different samples. By comparing these functions to those calculated for a Poisson point process known as complete spatial randomness (CSR) with the same average intensity (roughly, the number of points per volume), we can effectively identify and analyse spatial patterns. 

In our case, the nearest neighbour distances measure the separation between two events from the data set (here, the position of halos in the halo catalogue) or the distance between an event and a reference point (in our analysis, the reference points are randomly generated points in the simulation box with the same intensity as the halo catalogue).
The nearest neighbour
CDF, $G(r)$ indicates the proportion of the halos shown by $i$ having their first nearest neighbour on a sphere of radius $ r_i \le r $ around them. On the other hand, the spherical contact CDF, $F(r)$ shows the fraction of the reference points indicated by $j$ with an event point (position of a halo) within the distance $ {{r}}_j \le r $. 
The presence of clustering
leads to higher values for $ G(r) $ and $ F(r) $ in comparison to CSR \citep{hand2008statistical}. Moreover, \cite{Fard:2021qaa} argued that $ G(r) $ is advantageous for studying the high-density regions and small scales while $ F(r) $ comes in handy to investigate the larger scales. 
\\
An interpretable combination of these two functions is the $J-$function:
\begin{equation} \label{Eq-Jfun}
\\
J(r) \equiv \frac{1 - G(r)}{1 - F(r)},
\end{equation}
with the value one for CSR while in the presence of clustering, this function is less than unity. \\
The letter functions (i.e., $ G $, $ F $, and $ J $) are the statistics we use in this work. In the context of cosmology, whenever dealing with new statistics, two natural questions always arise: how much information these statistics contain and 
what is the connection between these statistics and the classic n-point correlation functions.
To address these, \cite{White:1979kp} introduced conditional correlation functions $ \Xi_i $, 
\begin{align} \label{Eq: conditional corr}
\Xi_i(\textbf{r}_1,...,\textbf{r}_i;V) =& \sum_{j=0}^{\infty} \dfrac{(-n)^j}{j!} \\ \nonumber
& \int ...\int \xi_{i+j}(\textbf{r}_1,...,\textbf{r}_{i+j})\,dV_{i+1}...dV_{i+j},
\end{align} 
where the integrals are over an empty volume of interest $V$, except for the points $ r_i $ and  $ \xi_i $ are the n-point correlation functions. Note that by definition $\xi_0 = 0$ and $\xi_1 = 1$.
\cite{White:1979kp} showed the probability of finding an empty sphere of radius $ r $ around a point, i.e., the void probability - $ P_0(V)$, is related to $ \Xi_0 $ through $ P_0(V) = \exp\left[ \Xi_0(V(r))\right] $, which is just the complement of $ F(r) $.
The probability of a halo having its first nearest neighbour on the surface or inside of a sphere of volume $ V $ is the complement of no other halo at this distance, which reads $ G(r) = 1 - \Xi_1(\textbf{r}_o; V(r)) \exp\left[ \Xi_0(V(r))\right] $.
\\
These relations, along with the equation \ref{Eq: conditional corr}, manifest the dependence of these statistics on not only the 2-point correlation function but on all the higher-order correlations.
Accordingly, it is reasonable to use these letter functions to investigate cosmological models, especially in non-linear regime. \\

\subsection{Massive neutrinos in cosmology}
The detection of the atmospheric neutrino oscillation indicates that at least two of three neutrino flavors of the standard model of particle physics have mass. The latest data release from  KATRIN experiment \citep{KATRIN:2021uub} suggests that the upper limit for the total neutrino mass would be around $\sum m_{\nu} \lesssim 0.8 \ $eV ($90 \% $ C. L.). 
For cosmology, this experimental fact means neutrinos belong to the matter sector rather than the radiation in late times. Consequently, the neutrino energy density at redshift zero reads,   
\begin{equation}
\\
\omega_{\nu} = \frac{ \sum m_{\nu} }{ 94.13 \rm  eV} \ ,  \label{eq:neutrino_density}
\end{equation}
with the neutrino mass fraction $f_{\nu} = {\omega_{\nu}}/{\omega_m} $, where $\omega_m=\Omega_m h^2$.
Hence, cosmological data sets can constrain the total neutrino mass through the constraints on the $ \omega_{\nu}$ independently from the laboratory kinematics constraints. 
This amount would be around $ \sum m_{\nu} \lesssim 0.12 \ $eV \citep{Planck:2018vyg}, with severe restrictions on the nature of the dark energy to be near the cosmological constant \citep{Hannestad:2005gj}\footnote{ If the dark energy equation-of-state and its derivative change concurrently, then the upper limit increases by a factor of three.}. 
Unlike cold dark matter particles, neutrinos fall out of equilibrium as relativistic particles. Due to the Universe's expansion, neutrinos lose energy and eventually become non-relativistic. This process introduces a non-relativistic length scale $ \lambda_{nr} $ as an ultimate scale beyond which neutrinos act like cold dark matter. Below this maximum scale, neutrinos stream freely and, due to their thermal velocities, can not cluster effectively under the influence of gravity. Accordingly, for a given redshift and neutrino species, the free-streaming wave number is defined as \cite{agarwal2011effect}:
\begin{equation}
\\
k_{f_s} (a) \approx 0.063 \, h {\rm Mpc}^{-1} \, \frac{m_{\nu}}{0.1 {\rm eV}} \frac{a^2 H(a)}{H_0}\ , 
\end{equation} 
with $ m_{\nu}$ is the mass of one neutrino specie,  $ a $ scale factor, $ H(a) $ the Hubble parameter, and $ H_0 = 100 h $ (km $ \rm {Mpc}^{-1} $ $ \rm s^{-1}$)
 the Hubble constant.
Moreover, free-streaming of the neutrinos smooth the matter density fluctuation field below $ \lambda_{fs} \sim k_{f_s}^{-1}$. This process leads to less clustering in the Universe and suppression of structures compared to the $ \Lambda$CDM universe.
\cite{LoVerde:2014rxa} shows that the presence of massive neutrinos reduces the abundance of massive halos, hosting galaxy clusters, by slowing down the collapse time. Non-linear clustering of massive neutrinos counter acts this delay slightly, but the outcome is the reduced number of massive halos by $ 1 \% $ for the total neutrino mass below $ 0.5 \ $eV.
Yet, there is a claim that the free-streaming of light relics induces a cut-off on the minimum mass of halos \cite{Schneider:2013ria}. The lighter the relic particles are, the cut-off in the halo mass gets smaller. \\
One of the main cosmological observations to probe the effect of massive neutrinos is the matter power spectrum both in linear and non-linear regimes \citep{Hannestad:2020rzl}.
To study the non-linear regime, one often uses the halo model approach. In this context, \cite{Hannestad:2020rzl} shows the suppression in power starts at $ k\simeq 0.01 \, h \ \rm{Mpc}^{-1} $, decreases up to $k\simeq 1 \, h \ \rm{Mpc}^{-1}$, and continues in larger wavenumbers as a plateau.
\cite{Hannestad:2020rzl} refer to the former feature as a "slide" and the latter as a "spoon" feature and state the spoon feature as a generic behaviour detected in different simulations.  As an example, \cite{Rossi:2017vmw}  used the simulation on the cosmic web and used the Lyman alpha observations to study the effect of neutrinos on small scales. 
Also, there are studies on the effect of massive neutrinos on void statistics and distribution \citep{Massara:2015msa}, also on the bias parameter {\citep{Hassani:2022yuq}}. 
\\
In this work, instead of using the 
power spectrum, we use the 1-point statistics to investigate the distribution of matter in the presence of massive neutrinos. We will show the probable relation of some specific scales, introduced in the two frameworks (power spectrum and 1-point statistics).

\section{Simulation and sample preparation} \label{Sec3}

In the following subsections, we discuss the simulation used and the data preparation process.
\subsection{Simulation}
\begin{figure}
\centering
\includegraphics[width=0.48\textwidth]{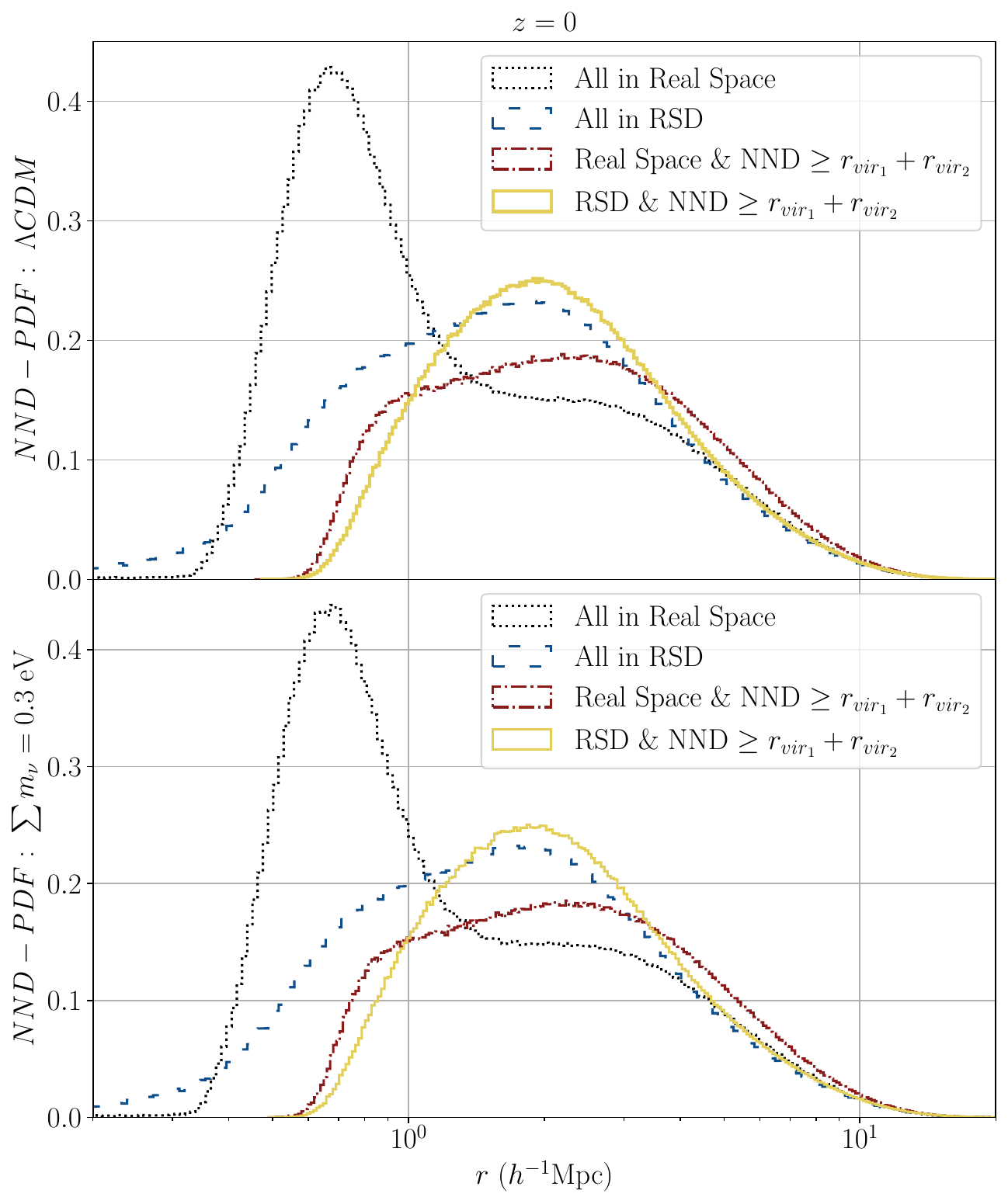}
\caption{The nearest-neighbour distance PDF for $\Lambda$CDM model (top) and $\sum m_{\nu} = 0.3 \ $ eV (bottom). The black dotted lines represent PDF in comoving space, the red dash-dotted lines show it after removing the intersecting halos, the blue dashed lines depict the PDF in the redshift space, and the solid yellow lines signify it after omitting the remaining intersecting halos in the redshift space. The x-axis represents the distance in real/redshift space, depending on whether the analysis is performed in either space.} 
\label{fig1-Bimodal-PDF}
\end{figure}

To study the effect of neutrino mass on the statistics introduced in Section \ref{Sec2}, we use gevolution code \citep{Adamek:2016zes, Adamek:2015eda}
\footnote{\url{https://github.com/gevolution-code}}
. Gevolution is a relativistic N-body code based on the weak field expansion of the general relativity. In this code, one can treat massive neutrinos as separate relativistic species or by using their linear perturbations from the transfer functions obtained from CLASS \citep{Blas:2011rf}. \cite{Adamek:2017uiq} showed that the results of the two approaches are consistent. In this work we use the halo catalogues of the simulations presented in \cite{Adamek:2017uiq}. Specifically, we use three simulations considering the total neutrino mass of  $ \sum\  m_{\nu} = 0.06 , \ 0.20, \  0.30 \ $ eV plus one for the vanilla $\Lambda$CDM cosmology. The fiducial cosmology parameters are given in Table \ref{Table1}. 
For the cosmologies with massive neutrinos, we set the density parameters of the cold dark matter according to the following relation, $\omega_c \equiv \Omega_c h^2= 0.12038 - \omega_{\nu}$, to keep the total  matter density ($\omega_{\nu}+\omega_{b}+\omega_{c}$) unchanged. In the previous relation  $\omega_{\nu}$ is computed using Eq.~\ref{eq:neutrino_density} and $\omega_c = 0.12038$ is the value we consider for the cold dark matter density in the $\Lambda$CDM scenario.
These simulations contain $4096^3$ particles and grids in a cube of the comoving box size $ L= 2048$ $h^{-1} {\rm Mpc}$ initialised at redshift $z=100$. 
Using these high resolution simulations we study the letter functions at $z=0$. Moreover, to study the redshift evolution of the 1-point statistics, we run lower resolution simulations, where we consider $L= 300$ $h^{-1} {\rm Mpc}$ and $N_{\rm grids} = N_{\rm pcl} = 1024^3$. We output snapshots at three redshifts  $z= 0, 0.5, 1$. The results for these simulations are presented in Appendix \ref{App_2}. It is worth mentioning that we construct the halo catalogues for each snapshot using ROCKSTAR halo finder \citep{behroozi2012rockstar} at  each redshift.

\begin{table}
\begin{center}
\begin{tabular}{ c | c | c | c | c | c }
$\omega_c$ & $\omega_b$ & $h$ & $n_s$ & $T_{\rm CMB}$ &  $ A_s $ \\ 
\hline \hline
$0.12038$ & $0.022032$ & $0.67556$ & $0.9619$ & $ 2.7255$ & $2.215 \times 10^{-9}$ \\ 
\end{tabular}
\caption{The cosmology parameters of standard model where $\omega_c=\Omega_ch^2$, $\omega_b=\Omega_b h^2$ are baryonic and dark matter densities normalises to critical density. $h=H_0 / 100$, $n_s$, $T_{\rm CMB}$, $ A_s$ are respectively the reduced Hubble parameter, spectral index, the CMB temperature, and the amplitude amplitude of scalar perturbations.}
\label{Table1}
\end{center}
\end{table}


\subsection{Data Preparation}
Data preparation follows these steps. First, we remove all sub-halos from the halo catalogue. Then we eliminate the halos which do not meet the virial condition: " $ \text{T} \  / \ \text{U} \  < 2 $", where T and U stand for kinetic and potential energies, respectively. 
 As we use a specific mass cut for sample preparation, we omit the sub-halos to control the effect of substructures. In observational proposal this means that we use a volume limited sample for our analysis. The halos which do not meet the virial condition are considered as ones which are not gravitationally bound objects \citep{Bett:2006zy}.
At this stage, if we compute the nearest neighbour distance and plot its probability distribution function (PDF)---referred to as NND-PDF---against the comoving radius,  $ r $, we observe a bimodal behaviour in the PDF. This bimodality appears for all simulations regardless of neutrino mass, i.e., dashed black line in the top and bottom panel of Figure \ref{fig1-Bimodal-PDF}, for  $\Lambda$CDM and $ \sum m_{\nu} = 0.3 $ eV respectively. 
Since our analysis depends on the behaviour of probability functions, it is critical to understand whether it is something physical or just an artefact of the resolution of the simulation or the halo finder used.
Our study suggests that most halos that contribute to the first mode are intersecting halos—denoting that these halos overlap with each other or share common spatial volume. To illustrate this, we compare the distance between the neighbouring halos with the sum of their virial radii and we find that some of the neighbouring halos intersect. If we omit the intersecting halos from the sample and then compute the NND, we end up with the red dash-dotted line in Figure \ref{fig1-Bimodal-PDF}. 
As we discuss in Appendix \ref{App_1}, we observe that if we do the same computations in the redshift space, where we use the redshift space distances instead of the comoving ones, we obtain a smooth PDF with one peak, and most of the contribution from intersecting halos introduced by the method of halo finder will disappear which is presented with the solid yellow line in Figure \ref{fig1-Bimodal-PDF}. To do this, we use the plane-parallel approximation and reposition all the halos along one Cartesian axis of the simulation box using the relation between the comoving positions, $ \textbf{x}_c $, and the one in the redshift space $ \textbf{x}_r$
\begin{equation} \label{Eq-RSMapping}
 \textbf{x}_r = \textbf{x}_c+ \bigl( \frac{\textbf{v} \  \cdot \  \hat{\textbf{n}}}{aH} \bigr)  \ \hat{\textbf{n}}  
\end{equation}
with $ \hat{\textbf{n}} $ being the unit vector along the line of sight, $ a $ scale factor, and $ H(z) $ the Hubble parameter at redshift $z$. {We consider two catalogues in redshift space: one that takes into account all halos and another that is filtered to eliminate intersecting halos. Eventually, the halo catalogues in the redshift space represent a smooth NND-PDF with one peak, the solid yellow and dashed blue lines in Figure \ref{fig1-Bimodal-PDF}. }
Hereafter, we do all the computations for the catalogues in the redshift space unless we state it directly. The results are almost independent of the specific direction we choose in simulation because of the statistical isotropy of the sample. Note that in the upcoming results the distances are in redshift space obtained from equation \ref{Eq-RSMapping}. 
 

\section{Results} \label{Sec4}
As mentioned earlier, to study the impact of neutrino mass, we use point process analysis tools to investigate the clustering of the halo catalogues. We specifically use the  
the so-called letter functions, i.e., $ G(r)-$ the nearest neighbour distance cumulative distribution function, $ F(r)- $the spherical contact cumulative distribution function, and $ J-$function as defined in equation \ref{Eq-Jfun}. 
The main task is to compare the letter functions for the base $ \Lambda$CDM and cosmologies with massive neutrinos ($\sum m_{\nu} = 0.06,0.2,0.3$ eV).
To do this, we compute the letter functions for each of the halo catalogues in different cosmologies. Then, we plot the subtraction of these functions for the catalogues with massive neutrinos and the $ \Lambda$CDM, i.e., $ \Delta_Y = Y_{\nu \Lambda {\rm CDM}} - Y_{\Lambda  {\rm CDM}}$, where Y is a member of the set $ \{G, F, J\} $.
\\
As G and F functions are CDFs, we subtract them in each radius for comparing two distinct {cosmological} models. This method is widely used in 
\cite{Banerjee:2020umh, Uhlemann:2019gni}.
We also compare the J-function in the same way to be consistent.
The errors, for each simulation and for each $Y$, are determined utilising the Jack-Knife method \citep{wolter2003introduction}. For this task, we divide the simulation box into $ n^3$ sub-boxes and subsequently omit each sub-box at the time, calculate the desired quantity, and at the end, perform averaging over the $ n^3-1$ quantities, and then using the mean as the data points and the standard deviation as the errors.  
The total error bars on the $\Delta_Y$ are calculated by the sum of the uncorrelated errors of the $ Y_{\Lambda  {\rm CDM}}$ and  $ Y_{\nu\Lambda  {\rm CDM}}$. 
The error bars are smaller than the size of the points.


\subsection{Letter function's dependence on the total neutrino mass}

\begin{figure}
\centering
\includegraphics[width=0.45\textwidth]{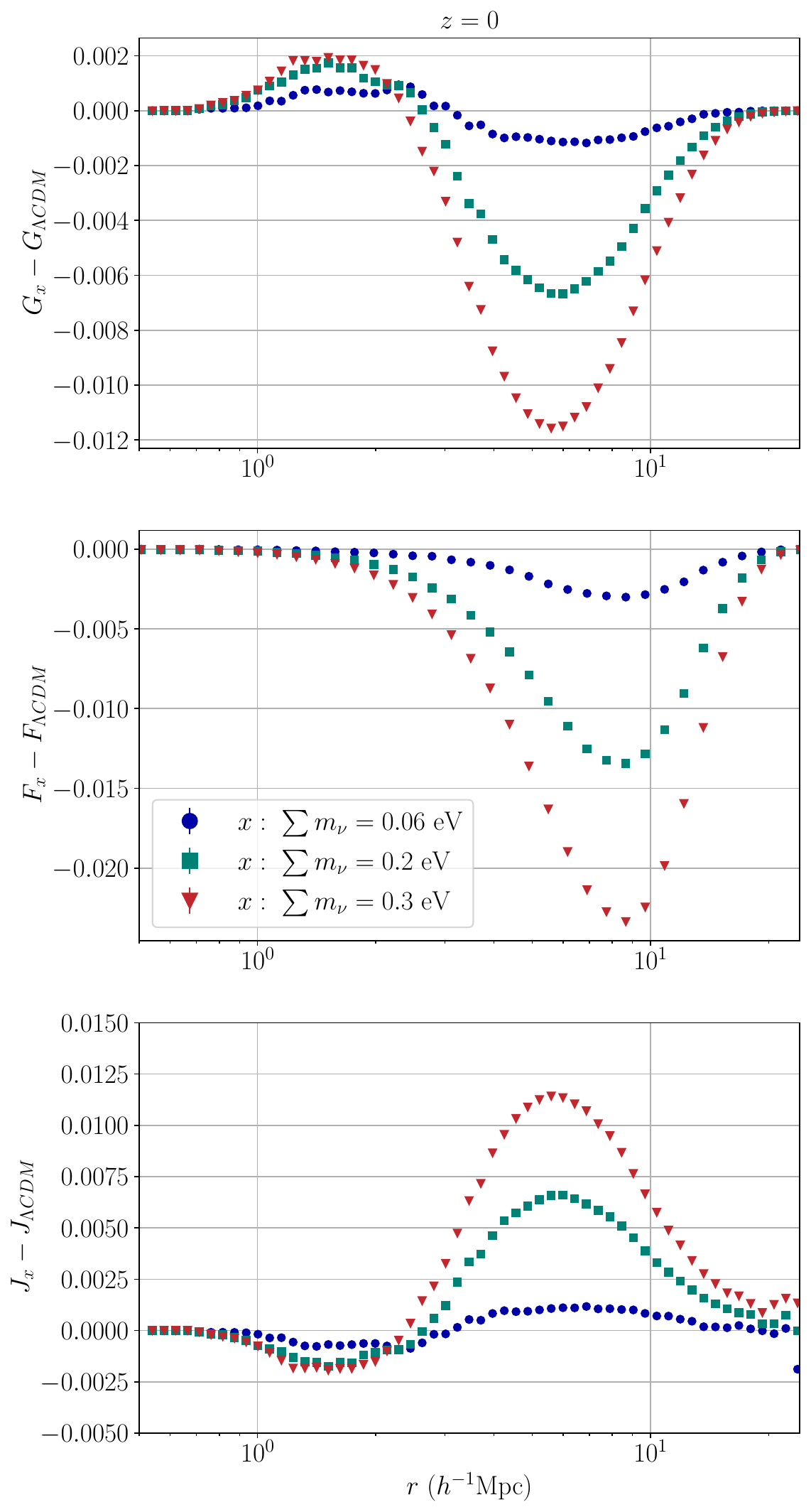}
\caption{The $\Delta_Y $ versus radius, blue circles related to $\sum m_{\nu} = 0.06 $ eV, green squares to $\sum m_{\nu} = 0.2 $ eV and red triangles to $\sum m_{\nu} = 0.3 $ eV. The top panel is $ \Delta_G$, the middle $ \Delta_F$, and the bottom $ \Delta_J $.}
\label{fig2-GFunc-Dif}
\end{figure}

In Figure \ref{fig2-GFunc-Dif}, we plot $ \Delta_Y $ versus distances in the redshift space at redshift $ z=0$. The blue circles, green squares, and red triangles are representative of the catalogue with the total neutrino mass of  0.06, 0.2, and 0.3  eV, respectively.
\\
The top panel depicts $ \Delta_G $, which manifests a generic behaviour, starting at zero, meeting a mild maximum value at around $ \sim 2 \ h^{-1} {\rm Mpc}$, then decreasing and hitting its minimum value around $\sim 7 \ h^{-1} {\rm Mpc}$, and finally reaching zero again. The depth of the $ \Delta_G$ is highly sensitive to the sum of the neutrino's mass. As the total mass increases, the $ \Delta_G$ deepens more. On the contrary, the peak of $\Delta_G$ is affected slightly by the change in the total mass. To interpret the behaviour, recall that $ G(r) $ is the probability of having the first nearest neighbour at distances less than $ r$ for the halos in the catalogue. Wherever the $\Delta_G$ takes a positive value, it implies finding the immediate neighbour is more probable in cosmology with massive neutrino than $\Lambda$CDM cosmology.  The negative value for $ \Delta_G$ translates to larger distances between neighbouring halos in neutrino cosmologies. 
Accordingly, the shape of the $ \Delta_G $ suggests that a small portion of halos in cosmology with massive neutrinos have their nearest neighbour at smaller distances than their counterpart in the base $ \Lambda$CDM, as for a large number of halos, the immediate neighbour is at further distances.
\\
{{To interpret Figure \ref{fig2-GFunc-Dif}, we relate the statistical properties of individual points to the behaviour of matter power spectrum when massive neutrinos are included in the matter sector. Typically, the main influence of massive neutrinos, attributed to their free streaming, is to dampen clustering at smaller scales and result in a decrease in matter power spectrum. However, we can explore the nonlinear regime of matter power spectrum using the halo model \citep{Cooray:2002dia}. The halo model suggests that, on small scales, the power spectrum is mainly affected by how matter is distributed within individual halos. On larger scales, within the quasi-linear regime, the power spectrum is derived from knowledge of both the matter distribution within halos and the density of dark matter halos. \cite{Hannestad:2020rzl} shows that in the context of the halo model, the power spectrum predicts a {\it{spoon-like}} shape for matter power spectrum in the presence of massive neutrinos. It means that the deviation from $\Lambda$CDM has its extremum at a scale of approximately $\sim 1 \ h^{-1} \rm{Mpc}$. One step further, in order to relate the matter power spectrum to the distribution of DM halos, one should account for the halo-bias term.  \cite{Hassani:2022yuq} investigate the scale-dependent nature of the halo bias term in the presence of massive neutrinos. In other words, their analysis illustrates that the bias parameter tends to be larger on smaller scales, which typically correspond to lighter halos. 
We argue that the peak at a smaller scale of $ \Delta_G $ arises from the one-halo term and the bias parameter at this scale. On the other hand, the minimum observed at a larger scale is a consequence of the free streaming effect, as a result of the dissipation of structures due to the presence of massive neutrinos.}} 
\\
In the middle panel, $ \Delta_F$ also reveals a generic behaviour that starts at zero, decreases, reaches a minimum at around  $\sim 10 \ h^{-1} {\rm Mpc}$, and then increases and goes to zero. Yet the total neutrino mass causes the change in the depth of the minimum value. As the sum of neutrinos' mass increases, the minimum value decreases. $F(r)$ 
 probes the clustering of the point process by measuring the distances between a set of randomly generated points and the halos from the catalogue. The shortfall of
 the probability for the neutrino cosmologies from their $\Lambda$CDM counterpart suggests that a randomly located sphere with a radius around $ \sim 10 \ h^{-1} {\rm Mpc}$ in a $\nu \Lambda$CDM is less populated than one placed in the $ \Lambda $CDM universe. It is worth mentioning that in general, the $F(r)$ is expected to probe larger scales than $G(r)$. This happens because in $G(r)$ both points are chosen from the halo samples while in $F(r)$ one of the points is chosen randomly.
\\
In the bottom panel of Figure \ref{fig2-GFunc-Dif}, we plot the $ \Delta_J $. 
The $ J$-function encapsulates the information in the $ G$ and $ F$ functions as defined in equation \ref{Eq-Jfun}.  The deviation of this function from unity toward lesser values represents clustering in the point pattern. The positive value for $ \Delta_J $ means that halos in $ \nu \Lambda$CDM cosmologies are less clustered than those halos in the $\Lambda$CDM one, and vice versa. The generic behaviour of $ \Delta_J $, inherited from $G$ and $F$ functions, once more suggests a slightly excessive clustering at a smaller scale and a fall-off of clustering on intermediate scales in cosmology with neutrinos to the base model. 
Note that the extremum on the $J$ and $G$ happens on the same scale, while the extremum on the $F$ is at a larger scale. This is because of the shape of the function of $J$ and the small correction of $F$ in the denominator causes this effect. An interesting point is that, in the matter power spectra, $\Lambda$CDM is always larger than massive neutrinos as neutrinos cannot cluster as effectively as CDM  particles. But we see a maximum in $\Delta G$ which seems to hint at a clustering at certain scales. But this is regarding the halos, not CDM particles. 
Although, the maximum is less significant in comparison to the minimum of the $\Delta_G$, however, it could be a specific fingerprint of the massive neutrinos. This fingerprint can be used to distinguish the massive neutrino from other types of the modified cold dark matter scenario, e.g., \cite{Kousha:2023kog}.
\\
These results propose that the notable signature of cosmology with massive neutrinos is in the dip of the $G$ and $F$ plots in the scales of $\sim 7-10 \ h^{-1} {\rm Mpc}$. 
The dip of the $G$ and $F$ is consistent with the less power in cosmology with massive neutrinos in the aforementioned scales. In order to propose an observational probe, we have to translate this distribution of dark matter halos to the galaxy distribution.
This translation is very complicated due to the halo occupation distribution and the bias parameter.

\subsection{Mass Cut }
\begin{figure}
\centering
\includegraphics[width=0.49\textwidth]{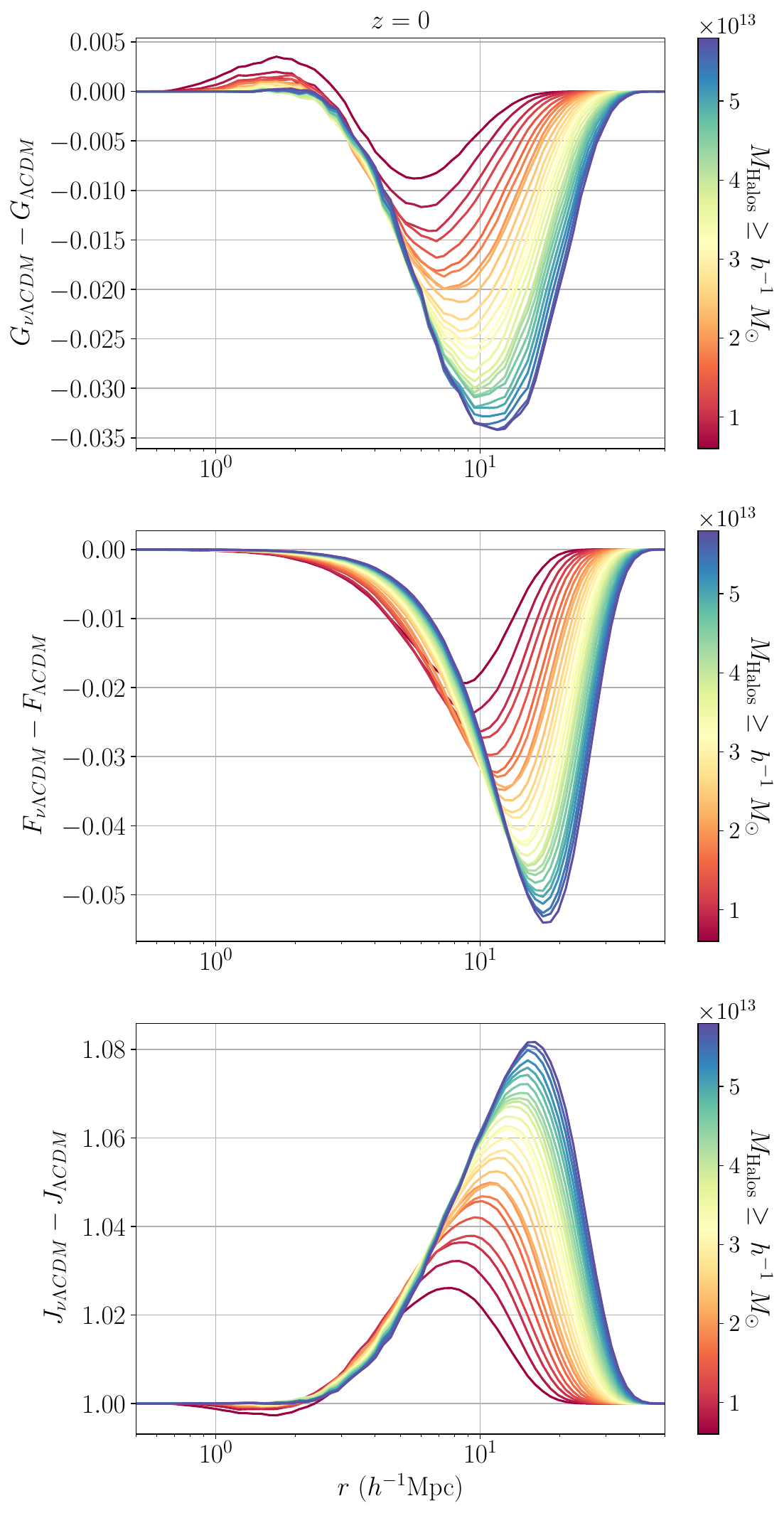}
\caption{The $\Delta_Y $ versus radius in the redshift space, for $ \sum m_\nu = 0.3 $ eV is plotted. The colour-bar shows the minimum mass of the sample.}
\label{fig:MassLim}
\end{figure}

In this subsection, at $z=0$, we investigate the dependency of mass cut on our results. The mass cut specifies the minimum mass for which we consider halos to be equal to or larger than it. By including a mass cut in our halo catalogues, we study various halo types that have undergone distinct dynamics. In Figure \ref{fig:MassLim}, we plot the $\Delta F$, $\Delta G$ and $\Delta J$ function versus the radius in the redshift space. The colour bar diagrams show the mass cut of these functions. By increasing the mass-cut
we observe that the minima in the $\Delta _{Y}$ where $Y=\{F,G\}$  deepen and occur at larger scales.
This behaviour is very similar to the behaviour of matter spectrum for different halo mass functions in small scales  \citep{Hannestad:2020rzl}. Higher mass cuts eliminate low-mass dark matter halos, which has the consequence of finding the nearest neighbours at larger distances. Among all letter functions the J-function shows the largest deviation from unity for large mass-cuts.

\section{Conclusions and Remarks} \label{Sec5}
Massive neutrinos affect the LSS mainly in non-linear regimes. Accordingly, the one point statistics which is a probe of the LSS in non-linear scales will be useful as a complementary method to traditional power spectrum calculations. In the data preparation, we follow the same procedure for the massive neutrino and $\Lambda$CDM cosmologies. We start the simulation from the same seed number and consider similar spatial resolution and halo finding processes.
This approach helps in mitigating discrepancies arising from simulations' randomness, numerics, or differences in spatial resolution. Consequently, the differences between the letter functions in the two cosmologies become more physical, minimising susceptibility to artificial effects introduced by the simulation.
In addition to the halo catalogue obtained from the halo finder, we prepare a halo catalogue in the redshift space. In this process, we transform the position of dark matter halos to the one in the redshift space observed by a distant observer and then we eliminate the halos that are in separation less than their total virial radius. In Appendix \ref{App_1} we show how this correction eliminates the artificial effect of halo halo-finding process. \\
We find the difference of the letter functions, namely G, F, and J functions, between the $\nu \Lambda$CDM and the base $\Lambda$CDM  cosmologies.
The specific shape of an excess in small scales and a deficit in larger scales is probably related to the characteristic scales appearing in the matter power spectrum.  In the context of the Halo model in smaller scales the one-halo term will be the dominant one and the two-halo terms of matter power spectrum affect the larger scales. The investigation of this relation is out of the scope of this work and is the subject of further study.\\
The redshift evolution of the letter functions is discussed
in the Appendix \ref{App_2}.
 We show that the difference between the two models occurs on almost the same scale with redshift evolution and the difference is more significant in higher redshifts. This suggests that high-$z$ surveys will be more convenient to distinguish the neutrino cosmology from $\Lambda$CDM using the letter functions. \\
The depth of the $\Delta_G$ and $\Delta_{F}$ is highly sensitive to the sum of neutrino masses; As the total neutrino mass increases, the $ \Delta_G$ deepens more. On the contrary, the peak of $\Delta_G$ is affected slightly by the change in the total neutrino mass.
\\
The extension of this work would be to investigate mock catalogues and study the distribution of galaxies in the light of 1-point statistics in both neutrino cosmology and standard model. 

\section*{Acknowledgments}
We would like to thank Negin Khosravaninezhad for many fruitful discussions. We are very grateful to Julian Adamek and Martin Kunz for giving us access to the halo catalogues of high resolution neutrino simulations that are presented in \citep{Adamek:2017uiq}. This work is supported by the University of Oslo
computational facilities and a grant from the Swiss National
Supercomputing Centre (CSCS) under project ID s1051. FH
is supported by the Overseas Research Fellowship from the
Research Council of Norway and the UNINETT Sigma2 -
the National Infrastructure for High Performance Computing
and Data Storage in Norway.
SB is partially supported by Abdus Salam International Center of Theoretical Physics (ICTP) under the regular associateship scheme.
Moreover, MK and SB are partially supported by the Sharif University of Technology Office of Vice President for Research under Grant No. G4010204.
MA is partially supported by Iran's National Elites Foundation (INEF) under the Chamran plan for young researchers. 
\section*{Data Availability}
The halo catalogue data, along with data for all plots derived from our theoretical models, can be made available upon request.




\bibliographystyle{mnras}
\bibliography{Bib.bib} 

\begin{thebibliography}{}
\makeatletter
\relax
\def\mn@urlcharsother{\let\do\@makeother \do\$\do\&\do\#\do\^\do\_\do\%\do\~}
\def\mn@doi{\begingroup\mn@urlcharsother \@ifnextchar [ {\mn@doi@}
  {\mn@doi@[]}}
\def\mn@doi@[#1]#2{\def\@tempa{#1}\ifx\@tempa\@empty \href
  {http://dx.doi.org/#2} {doi:#2}\else \href {http://dx.doi.org/#2} {#1}\fi
  \endgroup}
\def\mn@eprint#1#2{\mn@eprint@#1:#2::\@nil}
\def\mn@eprint@arXiv#1{\href {http://arxiv.org/abs/#1} {{\tt arXiv:#1}}}
\def\mn@eprint@dblp#1{\href {http://dblp.uni-trier.de/rec/bibtex/#1.xml}
  {dblp:#1}}
\def\mn@eprint@#1:#2:#3:#4\@nil{\def\@tempa {#1}\def\@tempb {#2}\def\@tempc
  {#3}\ifx \@tempc \@empty \let \@tempc \@tempb \let \@tempb \@tempa \fi \ifx
  \@tempb \@empty \def\@tempb {arXiv}\fi \@ifundefined
  {mn@eprint@\@tempb}{\@tempb:\@tempc}{\expandafter \expandafter \csname
  mn@eprint@\@tempb\endcsname \expandafter{\@tempc}}}

\bibitem[\protect\citeauthoryear{Adamek, Daverio, Durrer  \& Kunz}{Adamek
  et~al.}{2016a}]{Adamek:2016zes}
Adamek J.,  Daverio D.,  Durrer R.,   Kunz M.,  2016a, \mn@doi [JCAP]
  {10.1088/1475-7516/2016/07/053}, 07, 053

\bibitem[\protect\citeauthoryear{Adamek, Daverio, Durrer  \& Kunz}{Adamek
  et~al.}{2016b}]{Adamek:2015eda}
Adamek J.,  Daverio D.,  Durrer R.,   Kunz M.,  2016b, \mn@doi [Nature Physics]
  {10.1038/nphys3673}, 12, 346

\bibitem[\protect\citeauthoryear{Adamek, Durrer  \& Kunz}{Adamek
  et~al.}{2017}]{Adamek:2017uiq}
Adamek J.,  Durrer R.,   Kunz M.,  2017, \mn@doi [JCAP]
  {10.1088/1475-7516/2017/11/004}, 11, 004

\bibitem[\protect\citeauthoryear{Agarwal \& Feldman}{Agarwal \&
  Feldman}{2011}]{agarwal2011effect}
Agarwal S.,  Feldman H.~A.,  2011, Monthly Notices of the Royal Astronomical
  Society, 410, 1647

\bibitem[\protect\citeauthoryear{Aghamousa et~al.}{Aghamousa
  et~al.}{2016}]{DESI:2016fyo}
Aghamousa A.,  et~al., 2016, escholarship.org

\bibitem[\protect\citeauthoryear{Aghanim et~al.}{Aghanim
  et~al.}{2020}]{Planck:2018vyg}
Aghanim N.,  et~al., 2020, \mn@doi [Astron. Astrophys.]
  {10.1051/0004-6361/201833910}, 641, A6

\bibitem[\protect\citeauthoryear{Aker et~al.}{Aker
  et~al.}{2022}]{KATRIN:2021uub}
Aker M.,  et~al., 2022, \mn@doi [Nature Phys.] {10.1038/s41567-021-01463-1},
  18, 160

\bibitem[\protect\citeauthoryear{Amendola et~al.}{Amendola
  et~al.}{2018}]{Amendola:2016saw}
Amendola L.,  et~al., 2018, \mn@doi [Living Rev. Rel.]
  {10.1007/s41114-017-0010-3}, 21, 2

\bibitem[\protect\citeauthoryear{Banerjee \& Abel}{Banerjee \&
  Abel}{2020}]{Banerjee:2020umh}
Banerjee A.,  Abel T.,  2020, \mn@doi [Mon. Not. Roy. Astron. Soc.]
  {10.1093/mnras/staa3604}, 500, 5479

\bibitem[\protect\citeauthoryear{Banerjee, Kokron  \& Abel}{Banerjee
  et~al.}{2022}]{Banerjee:2021cmi}
Banerjee A.,  Kokron N.,   Abel T.,  2022, \mn@doi [Mon. Not. Roy. Astron.
  Soc.] {10.1093/mnras/stac193}, 511, 2765

\bibitem[\protect\citeauthoryear{Bardeen, Bond, Kaiser  \& Szalay}{Bardeen
  et~al.}{1986}]{Bardeen:1985tr}
Bardeen J.~M.,  Bond J.~R.,  Kaiser N.,   Szalay A.~S.,  1986, \mn@doi
  [Astrophys. J.] {10.1086/164143}, 304, 15

\bibitem[\protect\citeauthoryear{Behroozi, Wechsler  \& Wu}{Behroozi
  et~al.}{2012}]{behroozi2012rockstar}
Behroozi P.~S.,  Wechsler R.~H.,   Wu H.-Y.,  2012, The Astrophysical Journal,
  762, 109

\bibitem[\protect\citeauthoryear{Bett, Eke, Frenk, Jenkins, Helly  \&
  Navarro}{Bett et~al.}{2007}]{Bett:2006zy}
Bett P.,  Eke V.,  Frenk C.~S.,  Jenkins A.,  Helly J.,   Navarro J.,  2007,
  \mn@doi [Mon. Not. Roy. Astron. Soc.] {10.1111/j.1365-2966.2007.11432.x},
  376, 215

\bibitem[\protect\citeauthoryear{Blas, Lesgourgues  \& Tram}{Blas
  et~al.}{2011}]{Blas:2011rf}
Blas D.,  Lesgourgues J.,   Tram T.,  2011, \mn@doi [JCAP]
  {10.1088/1475-7516/2011/07/034}, 07, 034

\bibitem[\protect\citeauthoryear{Bond, Cole, Efstathiou  \& Kaiser}{Bond
  et~al.}{1991}]{Bond:1990iw}
Bond J.~R.,  Cole S.,  Efstathiou G.,   Kaiser N.,  1991, \mn@doi [Astrophys.
  J.] {10.1086/170520}, 379, 440

\bibitem[\protect\citeauthoryear{Bonnaire, Aghanim, Kuruvilla  \&
  Decelle}{Bonnaire et~al.}{2022}]{Bonnaire:2021sie}
Bonnaire T.,  Aghanim N.,  Kuruvilla J.,   Decelle A.,  2022, \mn@doi [Astron.
  Astrophys.] {10.1051/0004-6361/202142852}, 661, A146

\bibitem[\protect\citeauthoryear{Bonnaire, Kuruvilla, Aghanim  \&
  Decelle}{Bonnaire et~al.}{2023}]{Bonnaire:2022ocm}
Bonnaire T.,  Kuruvilla J.,  Aghanim N.,   Decelle A.,  2023, \mn@doi [Astron.
  Astrophys.] {10.1051/0004-6361/202245626}, 674, A150

\bibitem[\protect\citeauthoryear{Boyle, Uhlemann, Friedrich, Barthelemy, Codis,
  Bernardeau, Giocoli  \& Baldi}{Boyle et~al.}{2021}]{Boyle:2020bqn}
Boyle A.,  Uhlemann C.,  Friedrich O.,  Barthelemy A.,  Codis S.,  Bernardeau
  F.,  Giocoli C.,   Baldi M.,  2021, \mn@doi [Mon. Not. Roy. Astron. Soc.]
  {10.1093/mnras/stab1381}, 505, 2886

\bibitem[\protect\citeauthoryear{Cooray \& Sheth}{Cooray \&
  Sheth}{2002}]{Cooray:2002dia}
Cooray A.,  Sheth R.~K.,  2002, \mn@doi [Phys. Rept.]
  {10.1016/S0370-1573(02)00276-4}, 372, 1

\bibitem[\protect\citeauthoryear{Dodelson}{Dodelson}{2003}]{Dodelson:2003ft}
Dodelson S.,  2003, {Modern Cosmology}.
Academic Press, Amsterdam

\bibitem[\protect\citeauthoryear{Fard, Baghkhani, Ghodsi, Taamoli, Hassani  \&
  Baghram}{Fard et~al.}{2022}]{Fard:2021qaa}
Fard M.~A.,  Baghkhani Z.,  Ghodsi L.,  Taamoli S.,  Hassani F.,   Baghram S.,
  2022, \mn@doi [Mon. Not. Roy. Astron. Soc.] {10.1093/mnras/stac256}, 512,
  5165

\bibitem[\protect\citeauthoryear{Gough \& Uhlemann}{Gough \&
  Uhlemann}{2022}]{Gough:2021hlr}
Gough A.,  Uhlemann C.,  2022, \mn@doi [Universe] {10.3390/universe8010055}, 8,
  55

\bibitem[\protect\citeauthoryear{Hand}{Hand}{2008}]{hand2008statistical}
Hand D.~J.,  2008, Statistical analysis and modelling of spatial point patterns
  by janine illian, antti penttinen, helga stoyan, dietrich stoyan

\bibitem[\protect\citeauthoryear{Hannestad}{Hannestad}{2005}]{Hannestad:2005gj}
Hannestad S.,  2005, \mn@doi [Phys. Rev. Lett.]
  {10.1103/PhysRevLett.95.221301}, 95, 221301

\bibitem[\protect\citeauthoryear{Hannestad, Upadhye  \& Wong}{Hannestad
  et~al.}{2020}]{Hannestad:2020rzl}
Hannestad S.,  Upadhye A.,   Wong Y. Y.~Y.,  2020, \mn@doi [JCAP]
  {10.1088/1475-7516/2020/11/062}, 11, 062

\bibitem[\protect\citeauthoryear{Hassani, Adamek, Durrer  \& Kunz}{Hassani
  et~al.}{2022}]{Hassani:2022yuq}
Hassani F.,  Adamek J.,  Durrer R.,   Kunz M.,  2022, \mn@doi [Astron.
  Astrophys.] {10.1051/0004-6361/202244405}, 668, A56

\bibitem[\protect\citeauthoryear{Knebe et~al.}{Knebe
  et~al.}{2011}]{Knebe:2011rx}
Knebe A.,  et~al., 2011, \mn@doi [Mon. Not. Roy. Astron. Soc.]
  {10.1111/j.1365-2966.2011.18858.x}, 415, 2293

\bibitem[\protect\citeauthoryear{Kousha, Ansarifard  \& Abolhasani}{Kousha
  et~al.}{2023}]{Kousha:2023kog}
Kousha H.~M.,  Ansarifard M.,   Abolhasani A.,  2023,
  https://arxiv.org/abs/2312.10745

\bibitem[\protect\citeauthoryear{LoVerde}{LoVerde}{2014}]{LoVerde:2014rxa}
LoVerde M.,  2014, \mn@doi [Phys. Rev. D] {10.1103/PhysRevD.90.083518}, 90,
  083518

\bibitem[\protect\citeauthoryear{Massara, Villaescusa-Navarro, Viel  \&
  Sutter}{Massara et~al.}{2015}]{Massara:2015msa}
Massara E.,  Villaescusa-Navarro F.,  Viel M.,   Sutter P.~M.,  2015, \mn@doi
  [JCAP] {10.1088/1475-7516/2015/11/018}, 11, 018

\bibitem[\protect\citeauthoryear{Molin\'e et~al.}{Molin\'e
  et~al.}{2023}]{Moline:2021rza}
Molin\'e A.,  et~al., 2023, \mn@doi [Mon. Not. Roy. Astron. Soc.]
  {10.1093/mnras/stac2930}, 518, 157

\bibitem[\protect\citeauthoryear{Nikakhtar \& Baghram}{Nikakhtar \&
  Baghram}{2017}]{Nikakhtar:2016bju}
Nikakhtar F.,  Baghram S.,  2017, \mn@doi [Phys. Rev. D]
  {10.1103/PhysRevD.96.043524}, 96, 043524

\bibitem[\protect\citeauthoryear{Nikakhtar, Ayromlou, Baghram, Rahvar,
  Rahimi~Tabar  \& Sheth}{Nikakhtar et~al.}{2018}]{Nikakhtar:2018qqg}
Nikakhtar F.,  Ayromlou M.,  Baghram S.,  Rahvar S.,  Rahimi~Tabar M.~R.,
  Sheth R.~K.,  2018, \mn@doi [Mon. Not. Roy. Astron. Soc.]
  {10.1093/mnras/sty1415}, 478, 5296

\bibitem[\protect\citeauthoryear{Romanello et~al.}{Romanello
  et~al.}{2023}]{Romanello:2023obk}
Romanello M.,  et~al., 2023, https://arxiv.org/abs/2310.12224

\bibitem[\protect\citeauthoryear{Rossi}{Rossi}{2017}]{Rossi:2017vmw}
Rossi G.,  2017, \mn@doi [Astrophys. J. Suppl.] {10.3847/1538-4365/aa93d6},
  233, 12

\bibitem[\protect\citeauthoryear{Schneider, Smith  \& Reed}{Schneider
  et~al.}{2013}]{Schneider:2013ria}
Schneider A.,  Smith R.~E.,   Reed D.,  2013, \mn@doi [Mon. Not. Roy. Astron.
  Soc.] {10.1093/mnras/stt829}, 433, 1573

\bibitem[\protect\citeauthoryear{Sefusatti, Crocce, Scoccimarro  \&
  Couchman}{Sefusatti et~al.}{2016}]{Sefusatti:2015aex}
Sefusatti E.,  Crocce M.,  Scoccimarro R.,   Couchman H.,  2016, \mn@doi [Mon.
  Not. Roy. Astron. Soc.] {10.1093/mnras/stw1229}, 460, 3624

\bibitem[\protect\citeauthoryear{Sheth \& Tormen}{Sheth \&
  Tormen}{1999}]{Sheth:1999mn}
Sheth R.~K.,  Tormen G.,  1999, \mn@doi [Mon. Not. Roy. Astron. Soc.]
  {10.1046/j.1365-8711.1999.02692.x}, 308, 119

\bibitem[\protect\citeauthoryear{Sheth, Mo  \& Tormen}{Sheth
  et~al.}{2001}]{Sheth:1999su}
Sheth R.~K.,  Mo H.~J.,   Tormen G.,  2001, \mn@doi [Mon. Not. Roy. Astron.
  Soc.] {10.1046/j.1365-8711.2001.04006.x}, 323, 1

\bibitem[\protect\citeauthoryear{Stoyan}{Stoyan}{1984}]{stoyan1984correlations}
Stoyan D.,  1984, Mathematische Nachrichten, 116, 197

\bibitem[\protect\citeauthoryear{Takada, Komatsu  \& Futamase}{Takada
  et~al.}{2006}]{Takada:2005si}
Takada M.,  Komatsu E.,   Futamase T.,  2006, \mn@doi [Phys. Rev. D]
  {10.1103/PhysRevD.73.083520}, 73, 083520

\bibitem[\protect\citeauthoryear{Uhlemann, Friedrich, Villaescusa-Navarro,
  Banerjee  \& Codis}{Uhlemann et~al.}{2020}]{Uhlemann:2019gni}
Uhlemann C.,  Friedrich O.,  Villaescusa-Navarro F.,  Banerjee A.,   Codis S.,
  2020, \mn@doi [Mon. Not. Roy. Astron. Soc.] {10.1093/mnras/staa1155}, 495,
  4006

\bibitem[\protect\citeauthoryear{White}{White}{1979}]{White:1979kp}
White S. D.~M.,  1979, Mon. Not. Roy. Astron. Soc., 186, 145

\bibitem[\protect\citeauthoryear{Wolter \& Wolter}{Wolter \&
  Wolter}{2003}]{wolter2003introduction}
Wolter K.,  Wolter K.,  2003, Introduction to Variance Estimation.
Springer Series in Statistics, Springer, \url
  {https://books.google.nl/books?id=EadxTw0t2dMC}

\bibitem[\protect\citeauthoryear{Yuan, Zamora  \& Abel}{Yuan
  et~al.}{2023}]{Yuan:2023llf}
Yuan S.,  Zamora A.,   Abel T.,  2023, \mn@doi [Mon. Not. Roy. Astron. Soc.]
  {10.1093/mnras/stad1275}, 522, 3935

\bibitem[\protect\citeauthoryear{Zentner}{Zentner}{2007}]{Zentner:2006vw}
Zentner A.~R.,  2007, \mn@doi [Int. J. Mod. Phys. D]
  {10.1142/S0218271807010511}, 16, 763

\makeatother
\end{thebibliography}




\appendix
\section{Redshift evolution of 1-point statistics} \label{App_2}
To investigate the evolution of the letter functions (F, G, and J) in redshift, we run three additional simulations with a lower resolution including the total neutrino mass of  $ \sum m_{\nu} = 0, \  0.3, \ 0.6 \ $ eV using the gevolution code. 
In these simulations, we use the degenerate approximation for the mass of neutrino species, i.e., one massless and the two others having equal mass. Moreover, in gevolution, we consider the effect of massive neutrinos through their linear transfer functions provided by CLASS \footnote{\url{https://github.com/lesgourg/class_public}}
\citep{Blas:2011rf}.

The fiducial cosmology parameters are consistent with \cite{Planck:2018vyg} given in Table \ref{TableA1}. For the cosmologies with massive neutrinos, we set the density parameters of the dark matter according to the following relation, $\omega_c = 0.1200 - \sum m_{\nu}/ 94.13 \ $eV, to keep the total matter density unchanged. 

\begin{figure*}
\centering
\includegraphics[width=1\textwidth]{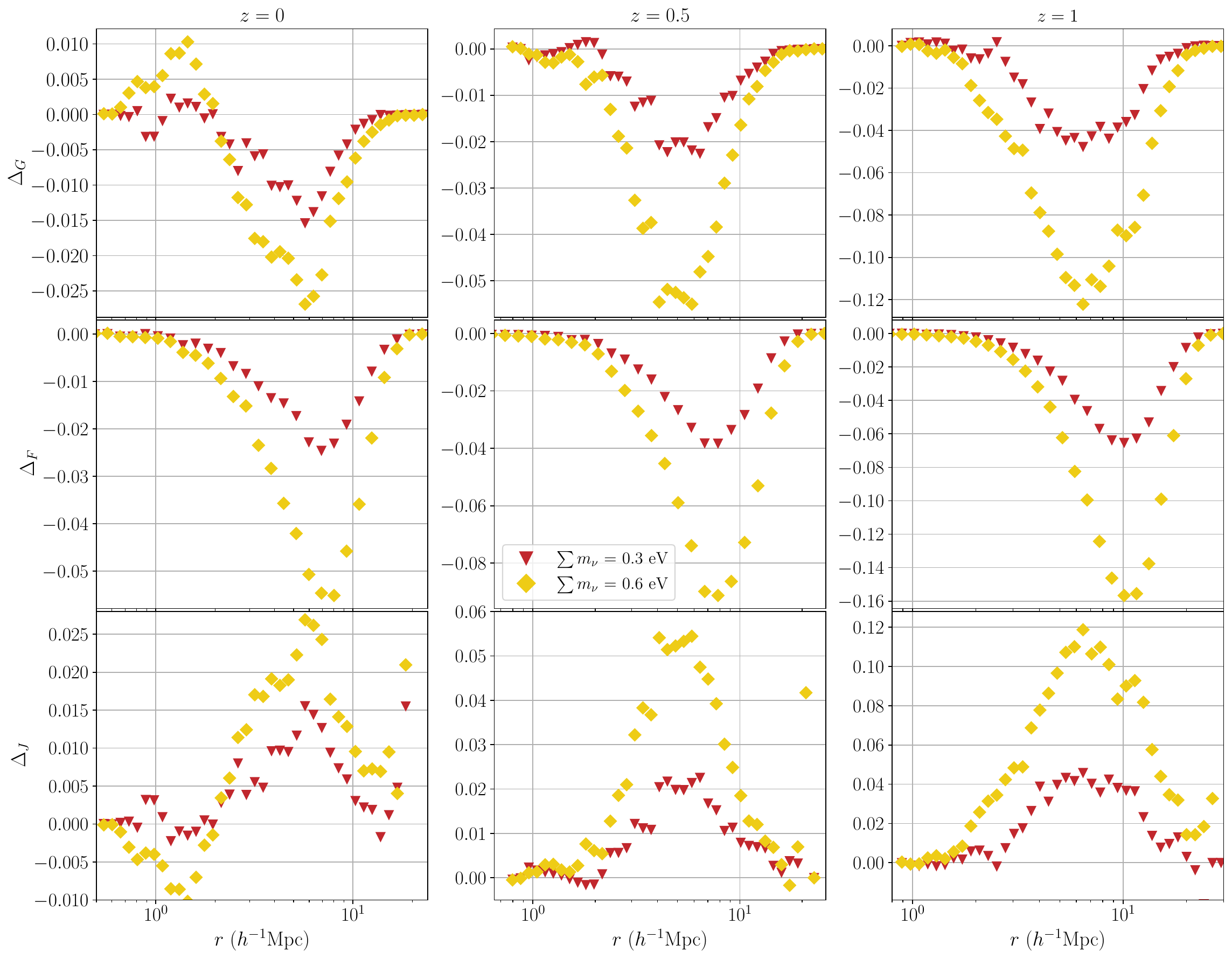}
\caption{We plot $\Delta_Y$ versus radius, where $Y=G,F,J$ for different redshifts $ z=0, 0.5, 1 $ and two total neutrino masses $ \sum m_{\nu} = 0.3, 0.6 $ eV. }
\label{fig3By3}
\end{figure*}

In these simulations, we consider the same number of particles and grid points $N_{\rm pcl} = N_{\rm grid}= 1024^3$ in a cube of the comoving length $ L = 300 \ h^{-1} {\rm Mpc}$  from $z_{\rm ini}=100$ to $z=0$. We output three snapshots at redshifts 0, 0.5, and 1. We construct the halo catalogues for each snapshot using the ROCKSTAR halo finder \citep{behroozi2012rockstar}.
The preparation of the datasets follows the steps discussed in Section \ref{Sec3} with an additional condition on the minimum mass of halos considered in all the snapshots. This mass reads  $ M_{min} = 5 \times 10^{12} \ h^{-1} M_\odot  $, and the idea behind it is closely connected to constructing a volume-limited galaxy catalogue.
\\
In Figure \eqref{fig3By3}, we plot the same letter functions as in Figure \ref{fig2-GFunc-Dif}, for redshifts $z=1, 0.5, 0$. We note that the local maximum on small scales disappears in higher redshifts in the $ G(r)$. However, the difference between the two models that are introduced in larger scales becomes more significant in higher redshifts. This means that surveys operating at higher redshifts are prominent to distinguish the neutrino cosmology models from the standard model. Our results are in the direction of the argument that high-$z$ surveys are more important than lower-$z$ surveys for neutrino mass detection \citep{Takada:2005si}.

\begin{table}
\begin{center}
\begin{tabular}{ c | c | c | c | c | c }
$\omega_c$ & $\omega_b$ & $h$ & $n_s$ & $\sigma_8$ &  $ln(10^{10} A_s)$ \\ 
\hline \hline
$0.12000$ & $0.02237$ & $0.6736$ & $0.9649$ & $ 0.8111$ & $3.044$ \\ 
\end{tabular}
\caption{The cosmology parameters of standard model where $\omega_c=\Omega_ch^2$, $\omega_b=\Omega_b h^2$ are baryonic and dark matter densities normalises to critical density. $h=H_0 / 100$,  $n_s$, $\sigma_8$ and $A_s$ are respectively the reduced Hubble parameter, spectral index, variance of perturbations in $8$ Mpc and the amplitude of scalar perturbations.}
\label{TableA1}
\end{center}
\end{table}

\section{Data preparation and simulation characteristics} \label{App_1}
\begin{figure}
\centering
\includegraphics[width=0.48\textwidth]{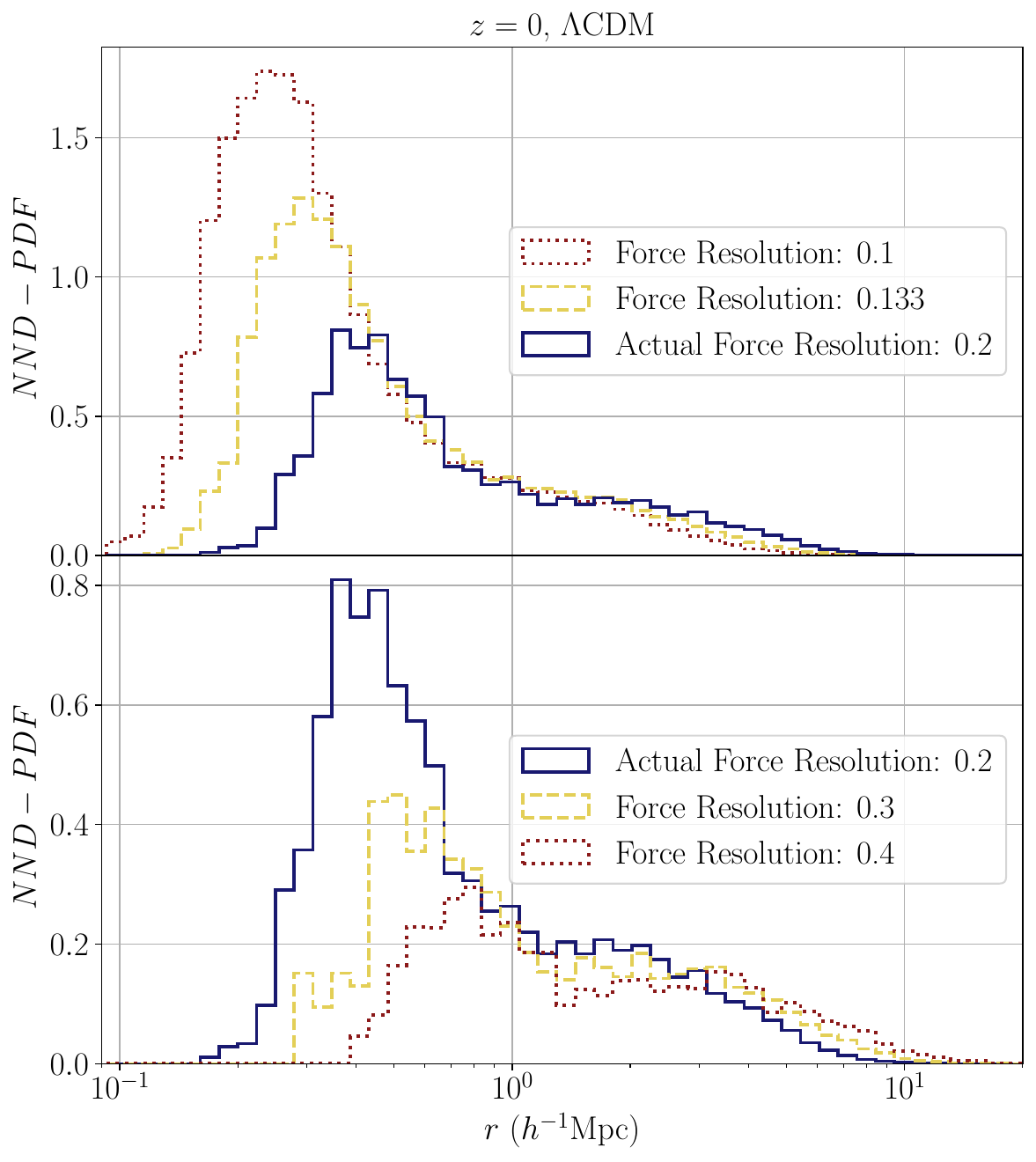}
\caption{The effect of the Force Resolution (FR) parameter on the peaks of the NND-PDF for one simulation.
The blue line in both the upper and lower panels corresponds to the actual force resolution. 
In the upper panel, the red dotted line and the dashed yellow line represent FR, which are half and two-thirds of the actual FR, respectively.
In the lower panel, the red dotted line and the dashed yellow line represent the FR, which are two times and 1.5 times the actual FR, respectively. We calculate PDFs in the coordinate space.}
\label{fig:app11}
\end{figure}

\begin{figure}
\centering
\includegraphics[width=0.48\textwidth]{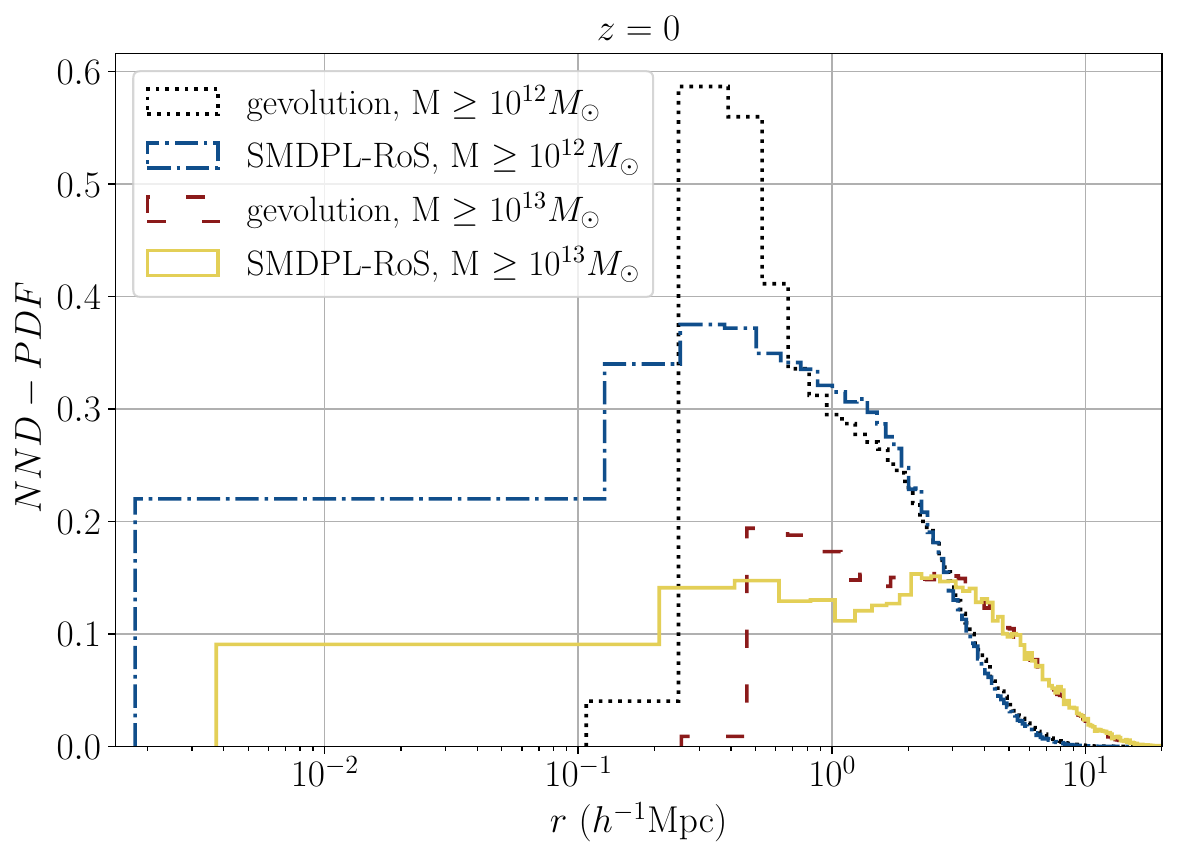}
\caption{The NND-PDF for gevolution and SMDPL in the coordinate space.}
\label{fig:app12}
\end{figure}

\begin{figure}
\centering
\includegraphics[width=0.48\textwidth]{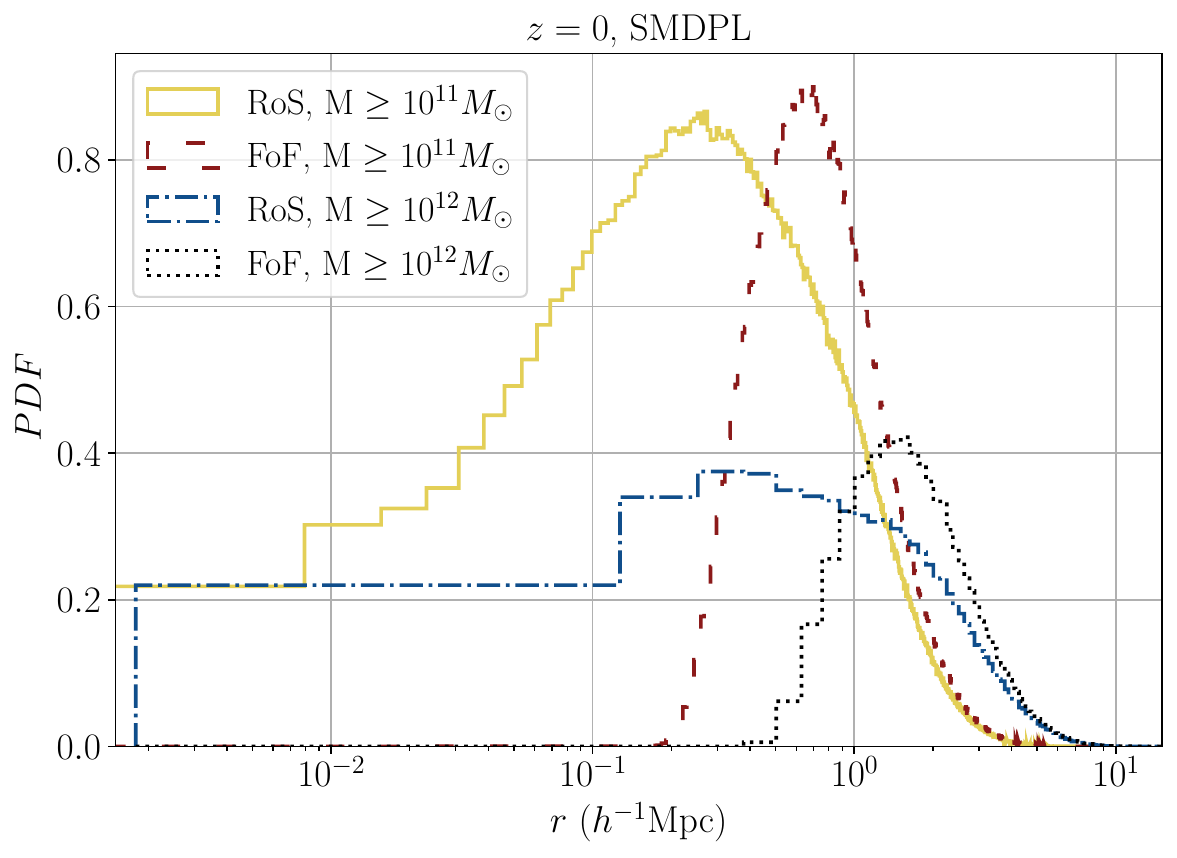}
\caption{
The NND-PDFs for SMDPL simulation and two different halo finder
algorithms versus distance in the coordinate space. 
The yellow solid and dashed red lines correspond to catalogues prepared by the ROCKSTAR (RoS) and Friend of Friend (FoF) when the minimum mass is set to $10^{11}  M_{\odot}$. 
The blue dash-dotted and dotted black lines relate to the RoS and FoF catalogues when the minimum mass is set to $10^{12}  M_{\odot}  $.   
}
\label{fig:app13}
\end{figure}

\begin{figure}
\centering
\includegraphics[width=0.48\textwidth]{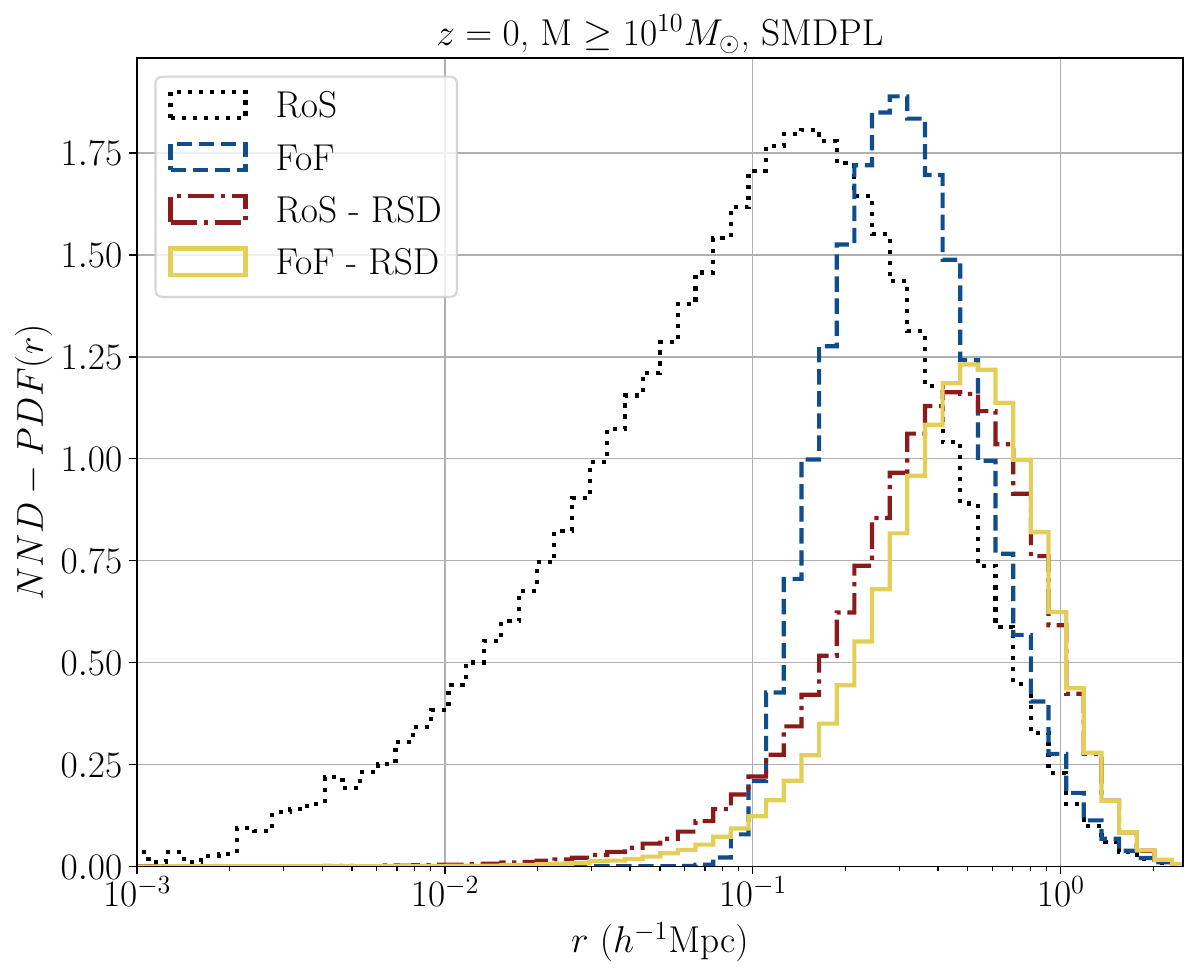}
\caption{The effect of the RSD transformation on the NND-PDF in SMDPL simulation for two halo finder algorithms.
The dotted black and dashed blue lines correspond to the real space distances for the RoS and FoF algorithms, respectively.
The dot-dashed red and solid yellow lines represent the NND-PDF after the RSD transformation of the distances for the RoS and FoF algorithms, respectively.}
\label{fig:app14}
\end{figure}

As mentioned in Section \ref{Sec3}, we observe two peaks in the NND-PDF of the halo catalogues from the gevelotion. By comparing the distance between two neighbours with the sum of their virial radii, we find that the intersecting halos contribute the most to the first peak. 
We see two possible roots for the situation. First, gevolution is a fixed-mesh simulation which results in a fixed spatial resolution and as a result the simulation might not be able to resolve the small scales properly. On the other hand, the ROCKSTAR performs the halo identification procedure in a 6+1 dimensional phase space. In this procedure, the velocity of the particles alters the spatial closeness, which is the main criterion used by the other halo-finder algorithms, such as Friend-of-Friend (FoF). To find out the extent of each factor on the peaks, we compare the NND-PDF
within different settings as follows:
\begin{itemize}
    \item \textbf{The effect of force resolution (FR) {on a single simulation}:} The force resolution which is a parameter in the halo finder ROCKSTAR, is related to the spatial resolution of the simulation. In our case, we consider $ FR \equiv (3/4\pi)^{1/3} (Ln^{-1}) $ where 
    $L$ is the box size and $n^3$ is the number of grids. In Figure \ref{fig:app11}, we plot the NND-PDF for different FR values in one snapshot from gevolution. It is evident from the plot that the first peak is ever-present, but it becomes less dominant as the FR increases. 
    \item \textbf{Comparison of two different particle mesh schemes:} We compare two halo catalogues, both prepared by ROCKSTAR, but from two distinct simulations, namely SMDPL \citep{Moline:2021rza}, and gevolution. To ensure the comparability of the two catalogues, we apply a mass cut and we compute the NND-PDF, which are shown in Figure \ref{fig:app12}. The dash-dotted blue and solid yellow lines correspond to the SMDPL simulation, while the dotted black and red dashed lines represent the gevolution simulation. In the case of the SMDPL, the higher spatial resolution of the simulation results in smaller initial distances between neighbouring halos, but it seems that the behaviour of the PDFs at large values is similar. Interestingly, the first peak of the NND-PDF also appears for the SMDPL when the mass cut is $ 10^{13}  M_{\odot}$.
    \item \textbf{The comparison between two halo finder:} The SMDPL simulation provides different halo catalogues, each prepared with a different algorithm. Figure \ref{fig:app13} shows the NND-PDFs for the ROCKSTAR (RoS in the figure) and the FoF catalogues for two different mass cuts. The FoF algorithm does the halo identification in the 3 dimensional coordinate space, and particle velocities do not play a role in the process. The figure shows that the PDFs for the ROCKSTAR catalogues have broader widths, and the peaks happen at smaller distances than for the FoF catalogues. 
    In the next part, we argue that the transformation to the redshift space will clean the effect of the halo finder algorithm in the halo identification process.
    \item \textbf{The impact of the transformation from coordinate space to the redshift space on the halo catalogues of the SMDPL simulation:} We use the ROCKSTAR and FoF halo catalogues from the SMDPL both with the mass cut of  $ 10^{10}  M_{\odot} $ and transform the position of the halos in both catalogues from the coordinate space to the redshift space using equation \ref{Eq-RSMapping} and using the parallel plane approximation. Figure \ref{fig:app14} shows the result in both the coordinate and the redshift spaces. Surprisingly, the two NND-PDF are alike in the redshift space, the dash-dotted red and solid yellow lines. 
\end{itemize}
{{Based on these observations and along with Figure \ref{fig1-Bimodal-PDF}, the method used for preparing the halo catalogues seems {important} to mitigate the unexpected effects rooted in the simulation resolution and halo-finding algorithms for the statistical measures used in the current paper. \\
Although, the plausibility of this conclusion needs further consideration. \cite{Knebe:2011rx} provide an extensive comparison of various halo-finding algorithms. As mentioned in this work, for the FoF algorithm, the building blocks---the clusters (then to be recognised as halos) reside in almost non-overlapping volumes in the coordinate space. Weighted graphs are used to construct clusters, and the identification process is based on the graph without loops (minimum spanning trees) between particles and their spatial proximity.
The ROCKSTAR algorithm also starts grouping the particles using a FoF scheme at large distances, then continues the halo identification process in the phase space and uses a modified metric to decide whether two particles belong to one halo. This metric for two particles with position and velocity of 
$(\bm{x}_1, \bm{v}_1)$ and $ (\bm{x}_2, \bm{v}_2)$  is defined as  $ d_{12} = \sqrt{ \frac{(\bm{x}_1 - \bm{x}_2)^2}{\sigma^2_x} + \frac{(\bm{v}_1 - \bm{v}_2)^2}{\sigma^2_v}}$ where $ \sigma_x $ and $ \sigma_v$
are the standard deviations for particle position and velocity of the subgroup respectively. The second term in the square is a positive quantity, and it can cause two close particles in the coordinate space to move apart in the phase space. This metric can result in identifying multiple halos for a set of particles considered as a single halo based on their spatial proximity in the coordinate space. It also leads to overlapping halos in the coordinate space, as opposed to halos identified by the FoF algorithm. }}
{{Next, we discuss the impact of transforming the halo positions to the redshift space using equation \ref{Eq-RSMapping}. As per Section \ref{Sec3}, we use the parallel plane approximation and adjust the $z$-component of positions by ${v_z}/({aH})$, where $v_z$ is the z-component of the velocity. In this process, a few percent of halos from both halo catalogues (FoF and ROCKSTAR from the SMDPL simulation) leave the simulation box \mkn{\footnote{They were placed back into the simulation box using the periodic boundary condition.}}, which indicates the stretching along the z-axis. Repositioning the halos in this way causes an increase in their distance, which shows itself in the overall shift of both NND-PDFs towards the large distances. This repositioning seems to counteract the effect of the phase space metric for the ROCKSTAR catalogues, and this causes the two NND-PDFs to become more similar to each other.}}
\\

\bsp	
\label{lastpage}
\end{document}